\documentclass[reprint, amsmath,amssymb,superscriptaddress,prb]{revtex4-2}

\usepackage{graphicx}
\usepackage{bm}
\usepackage{txfonts}
\usepackage{xspace}
\usepackage{color}

\newcommand{\EF}{$E_\mathrm{F}$\xspace}
\newcommand{\Thd}{$\mathrm{Th}~6d$\xspace}
\newcommand{\Uf}{$\mathrm{U}~5f$\xspace}

\newcommand{\Pdd}{$\mathrm{Pd}~4d$\xspace}

\newcommand{\Alp}{$\mathrm{Al}~3p$\xspace}

\newcommand{\orb}[2]{$\mathrm{ #1 } ~ #2 $\xspace}
\newcommand{\hn}[1]{$h\nu #1~\mathrm{eV}$\xspace}
\newcommand{\EB}[1]{$E_{\mathrm{B}} #1~\mathrm{eV}$\xspace}

\newcommand{\Gm}{$\mathrm{\Gamma}$\xspace}
\newcommand{\pnt}[1]{$\mathrm{#1}$\xspace}

\newcommand{\ThPdAl}{$\mathrm{ThPd_2Al_3}$\xspace}
\newcommand{\ThRuSi}{$\mathrm{ThRu_2Si_2}$\xspace}
\newcommand{\UPdAl}{$\mathrm{UPd_2Al_3}$\xspace}
\newcommand{\URuSi}{$\mathrm{URu_2Si_2}$\xspace}

\begin{document}

\preprint{APS/123-QED}

\title{Electronic Structure of $\mathrm{ThPd_2Al_3}$: an impact of the $\mathrm{U}~5f$ states in the electronic structure of $\mathrm{UPd_2Al_3}$}

\author{Shin-ichi~Fujimori}
\affiliation{Materials Sciences Research Center, Japan Atomic Energy Agency, Sayo, Hyogo 679-5148, Japan}

\author{Yukiharu~Takeda}
\affiliation{Materials Sciences Research Center, Japan Atomic Energy Agency, Sayo, Hyogo 679-5148, Japan}

\author{Hiroshi~Yamagami}
\affiliation{Materials Sciences Research Center, Japan Atomic Energy Agency, Sayo, Hyogo 679-5148, Japan}
\affiliation{Department of Physics, Faculty of Science, Kyoto Sangyo University, Kyoto 603-8555, Japan}

\author{Ji\v{r}\'i~Posp\'i\v{s}il}
\affiliation{Advanced Science Research Center, Japan Atomic Energy Agency, Tokai, Ibaraki 319-1195, Japan}
\affiliation{Department of Condensed Matter Physics, Faculty of Mathematics and Physics, Charles University,  
Ke Karlovu 5, 121 16 Prague 2, Czech Republic}

\author{Etsuji~Yamamoto}
\affiliation{Advanced Science Research Center, Japan Atomic Energy Agency, Tokai, Ibaraki 319-1195, Japan}

\author{Yoshinori~Haga}
\affiliation{Advanced Science Research Center, Japan Atomic Energy Agency, Tokai, Ibaraki 319-1195, Japan}

\date{\today}

\begin{abstract}
The electronic structure of $\mathrm{ThPd_2Al_3}$, which is isostructural to the heavy fermion superconductor $\mathrm{UPd_2 Al_3}$, was investigated by photoelectron spectroscopy.
The band structure and Fermi surfaces of $\mathrm{ThPd_2Al_3}$ were obtained by angle-resolved photoelectron spectroscopy (ARPES), and the results were well-explained by the band-structure calculation based on the local density approximation.
The comparison between the ARPES spectra and the band-structure calculation suggests that the Fermi surface of \ThPdAl mainly consists of the $\mathrm{Al}~3p$ and $\mathrm{Th}~6d$ states with a minor contribution from the $\mathrm{Pd}~4d$ states.
The comparison of the band structures between $\mathrm{ThPd_2Al_3}$ and $\mathrm{UPd_2Al_3}$ argues that the $\mathrm{U}~5f$ states form Fermi surfaces in $\mathrm{UPd_2Al_3}$ through hybridization with the $\mathrm{Al}~3p$ state in the $\mathrm{Al}$ layer, suggesting that the Fermi surface of $\mathrm{UPd_2Al_3}$ has a strong three-dimensional nature.
\end{abstract}

\maketitle

\section{Introduction}
The coexistence of unconventional superconductivity and antiferromagnetic ordering with relatively large magnetic moment (0.85~$\mu_{\mathrm{B}} / \mathrm{U}$) is the most characteristic feature of the heavy fermion compound \UPdAl \cite{UPd2Al3_Geibel}.
There are many electronic structure studies for \UPdAl; however, the nature of the \Uf state in \UPdAl remains contradictory.
The dHvA study reported that the observed dHvA branches are mostly explained by the band structure calculation treating the \Uf states as being itinerant \cite{UPd2Al3_dHvA}.
Its electronic structure has also been studied by photoelectron spectroscopy \cite{UPd2Al3_ARPES1, UM2Al3_ARPES, Ucore, UPd2Al3_ARPES2, SF_review_JPSJ}, and the itinerant nature of the \Uf states has been reported.
A recent Compton scattering study has also reported that the \Uf states have an itinerant character at temperatures lower than $T = 20~\mathrm{K}$ \cite{UPd2Al3_Compton}, which is consistent with the ARPES study \cite{UPd2Al3_ARPES1}.
Meanwhile, a resonant photoemission study on \UPdAl has also suggested the itinerant nature of the \Uf state, but there exists the correlated satellite structure due to the strong electron correlation effect \cite{U4d5fRPES}.
In contrast, the transport properties of \UPdAl are very similar to those of heavy fermion Ce-based compounds, and they have been essentially understood based on the very localized \Uf picture.
For example, the temperature dependence of the magnetic susceptibility has been interpreted based on the crystalline electric field scheme \cite{UPd2Al3_tetraU}.
To understand these physical properties consistently, the dual model of the \Uf states has been proposed for this compound \cite{UPd2Al3_Zwicknagl1}.
In this scenario, the \Uf states are formally divided into two subsystems: an itinerant $f^1$ component and a localized $f^2$ component.
The result of the recent point contact spectroscopy study on \UPdAl was also interpreted along with this scenario \cite{UPd2Al3_point}, but the microscopic information of the \Uf states is still lacking.
To understand the electronic structure of \UPdAl, it is essential to identify the contribution of the \Uf state to gain a greater insight regarding its nature in this compound.

In the present study, we have studied the electronic structure of \ThPdAl, which is the $f^0$-reference compound of \UPdAl, by photoelectron spectroscopy.
We compared the angle-resolved photoelectron spectroscopy (ARPES) spectra between \ThPdAl and \UPdAl, and the contribution from the \Uf states to the band structure of \UPdAl was clarified.
Furthermore, \ThPdAl is a superconductor with $T_\mathrm{C}=0.2~\mathrm{K}$ \cite{ThPd2Al3_1}.
It is the only case where an uranium heavy fermion superconductor has an isostructural counterpart of the thorium compound, which is also superconductor.
Thus, \ThPdAl is in itself an important target material to study its electronic structure.
We found that the band structure and the topology of the Fermi surface of \ThPdAl are essentially explained by the band-structure calculation based on the local density approximation.

\section{Experimental Procedure and Band-Structure Calculation}
The \ThPdAl single crystal was grown by Czochralski method by pulling speed of $12~\mathrm{mm/h}$ in a tetraarc furnace under Ar protective atmosphere.
Final shape of single crystal was a cylinder with a diameter of $2-3~\mathrm{mm}$ and a length of $15~\mathrm{mm}$.
The high quality of single crystal was confirmed by Laue method showing sharp spots and RRR ratio = 37 (the extrapolated value $\rho_0 = 0.7~\mathrm{\mu \Omega / cm}$ ) after 2 weeks annealing at $900\mathrm{^\circ C}$ in an evacuated quartz ampule.
Photoemission experiments were performed at the soft X-ray beamline BL23SU of SPring-8 \cite{BL23SU2}.
The overall energy resolution in angle-integrated photoemission (AIPES) experiments at \hn{=800} was approximately $150~\mathrm{meV}$, and that in the ARPES experiments at \hn{=660} was approximately $90~\mathrm{meV}$.
Clean sample surfaces were obtained by cleaving the samples {\it in situ} perpendicular to the $c$ axis under ultra-high vacuum conditions.
The vacuum during the course of measurements was typically $<2 \times 10^{-8}~\mathrm{Pa}$, and the sample surfaces were stable for the entire duration of the measurements (about 2 days) since no significant changes had been observed in the ARPES spectra during these periods.
The positions of the ARPES cuts in the momentum space were determined by assuming a free-electron final state with an inner potential of $V_{0}=12~\mathrm{eV}$.

In the band-structure calculations, relativistic linear augmented plane wave (RLAPW) calculations \cite{Yamagami} within the LDA \cite{LDA} were performed, treating all \Uf electrons as itinerant.
In this approach, the Dirac-type Kohn-Sham equation has been formulated, and the spin-orbit interaction is exactly taken into account \cite{suppl}.
To compare the results of the calculation with the ARPES spectra, we have simulated the ARPES spectral functions on the basis of the band-structure calculation.
In the simulation, the following effects were taken into account: (i) the broadening along the $k_\perp$ direction due to the finite escape depth of the photoelectrons, (ii) the lifetime broadening of the photohole, (iii) the photoemission cross sections of orbitals, and (iv) the energy resolution and the angular resolution of the electron analyzer.
The details are outlined in Ref.~\cite{UN_ARPES}.

\section{Results and Discussion}
\begin{figure}
	\centering
	\includegraphics[scale=0.5]{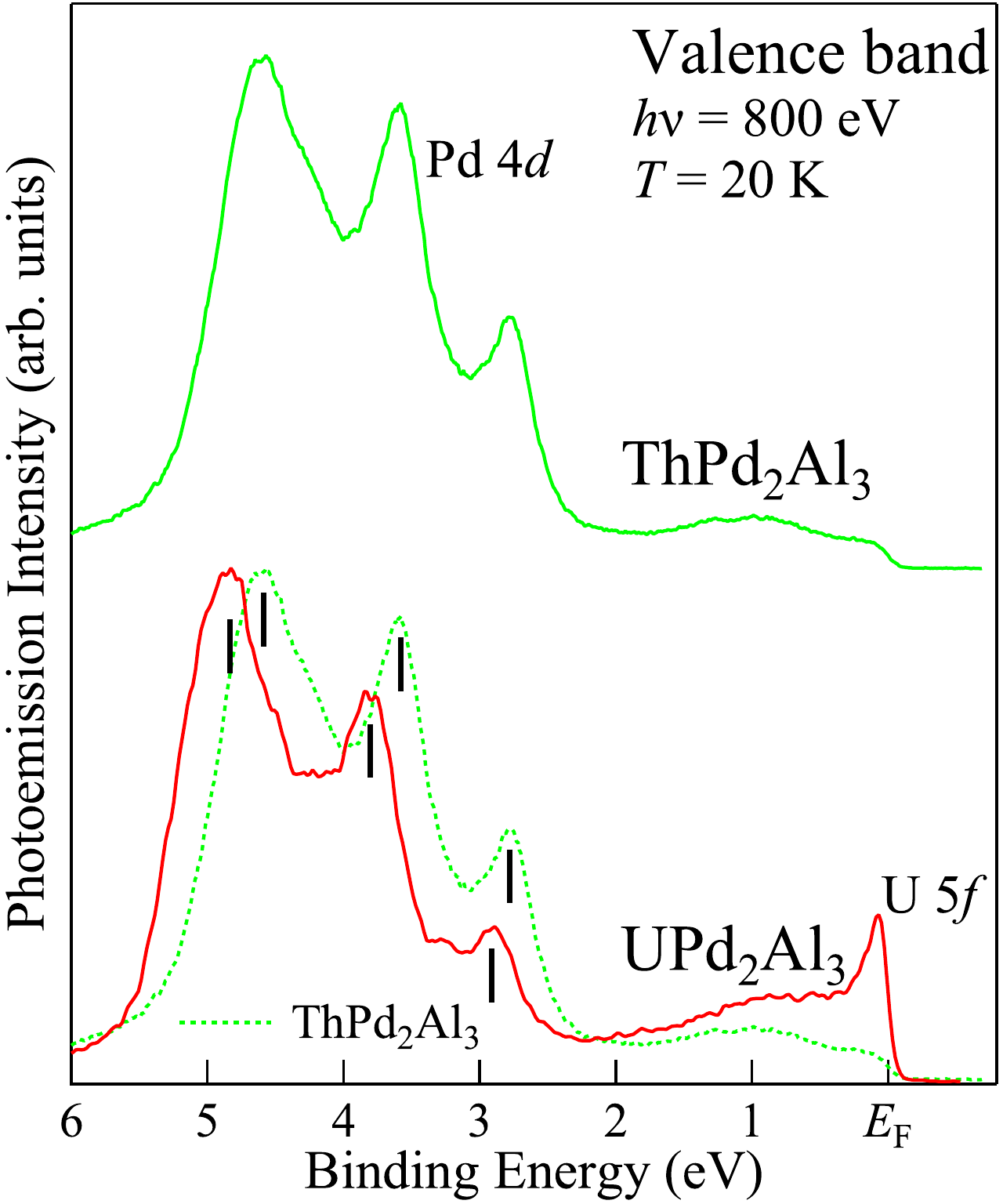}
	\caption{Valence-band spectrum of \ThPdAl measured at \hn{=800}. Also, the bottom part of this figure provides a comparison of the valence band spectra of \ThPdAl and \UPdAl measured at \hn{=800}.
}
	\label{val}
\end{figure}
\begin{figure*}[t]
	\centering
	\includegraphics[scale=0.5]{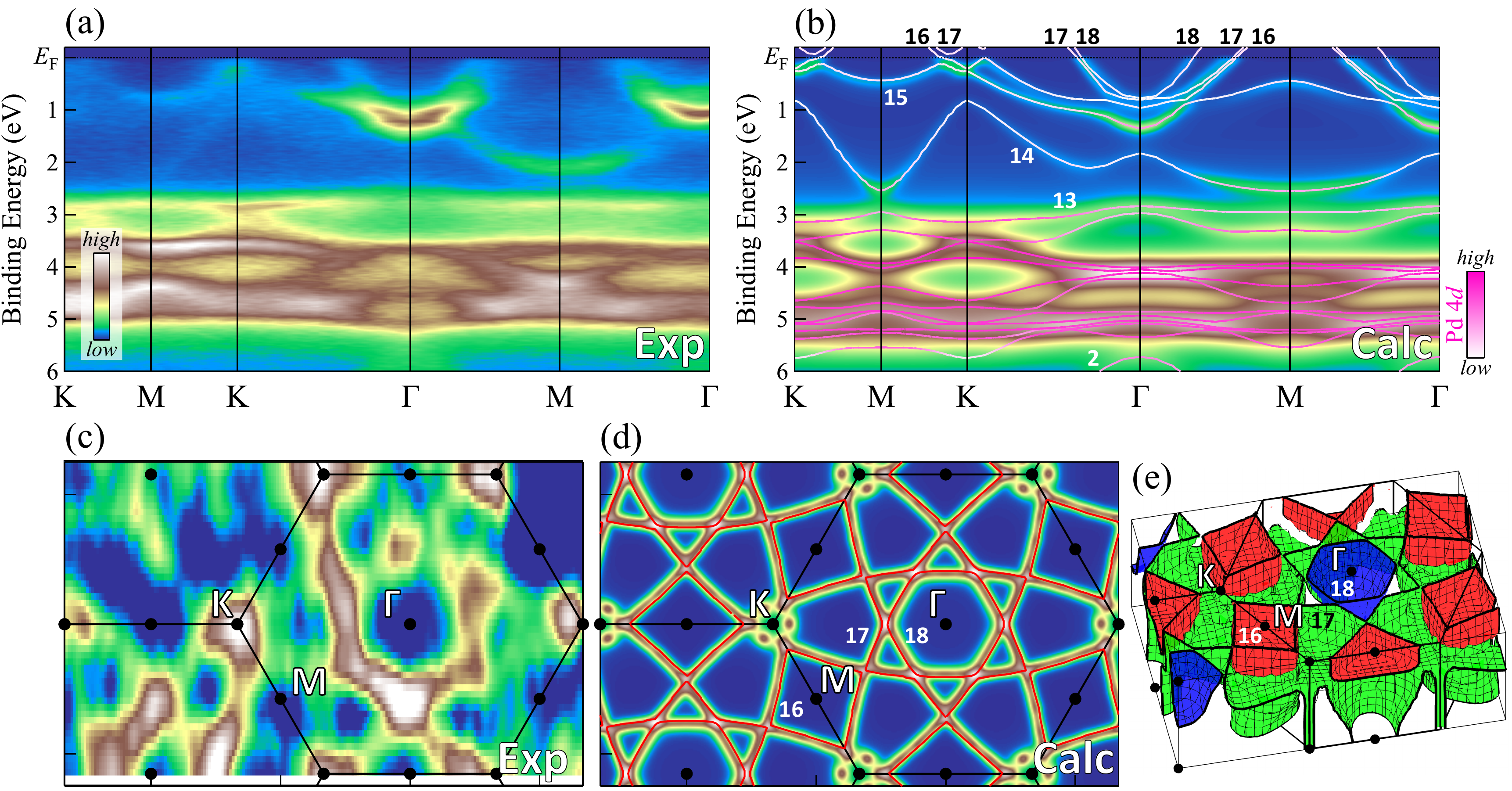}
	\caption{Band structure and Fermi surface of \ThPdAl obtained by ARPES.
	(a) ARPES spectra of \ThPdAl measured along the \pnt{K}-\pnt{M}-\pnt{K}-\Gm-\pnt{M}-\Gm high-symmetry line at \hn{=660}.
	(b) Calculated band structure and simulation of the ARPES spectra based on the band-structure calculation.
	(c) Fermi surface map obtained by the integration of the ARPES spectra over 100~meV across \EF.
	(d) Calculated Fermi surface and the simulation of the experimental Fermi surface map.
	(e) Three-dimensional shape of the calculated Fermi surfaces.
}
	\label{arpes}
\end{figure*}
Figure~\ref{val} shows the AIPES valence-band spectrum of \ThPdAl measured at \hn{=800} and its comparison with that of \UPdAl.
According to the photoionization cross-sections, the contributions from the \Pdd and \Uf states are dominant in this photon energy \cite{atomic}, and thus three prominent peaks located between \EB{=2.5-6} are ascribed to the contributions from the \Pdd states.
The bottom part of this figure demonstrates the comparison between the valence-band spectra of \ThPdAl and \UPdAl measured at \hn{=800}.
The comparison shows that the sharp peak at the Fermi energy exists only in the spectrum of \UPdAl, suggesting that the peak found at the Fermi energy represents the contribution of the \Uf states in \UPdAl.
This is consistent with the result of the resonant photoemission experiment for \UPdAl in which a similar similar sharp peak was observed at the Fermi energy of \UPdAl \cite{U4d5fRPES}.
A further important point to note is that all three \Pdd peaks in the valence-band spectrum of \UPdAl are located in higher binding energies than those in the valence-band spectrum of \ThPdAl.
A very similar rigid shift of the transition metal $d$ bands has been observed in the valence-band spectra of \URuSi and \ThRuSi \cite{ThRu2Si2_ARPES}.
The amount of energy shift is approximately $200~\mathrm{meV}$, which is also very similar to the case between \ThRuSi and \URuSi.
This energy shift of the $d$ band is in contrast with the case of the \Uf localized compound $\mathrm{UPd_3}$ where no shift was observed in the \Pdd bands between $\mathrm{ThPd_3}$ and $\mathrm{UPd_3}$ \cite{UThPd3}.
The rigid shift in the \Pdd states between \UPdAl and \ThPdAl suggests that the \Uf states are involved in the valence-band structure of \UPdAl, and they are not impurity-like contributions as expected when the \Uf states are almost localized.

\begin{figure*}[t]
	\centering
	\includegraphics[scale=0.5]{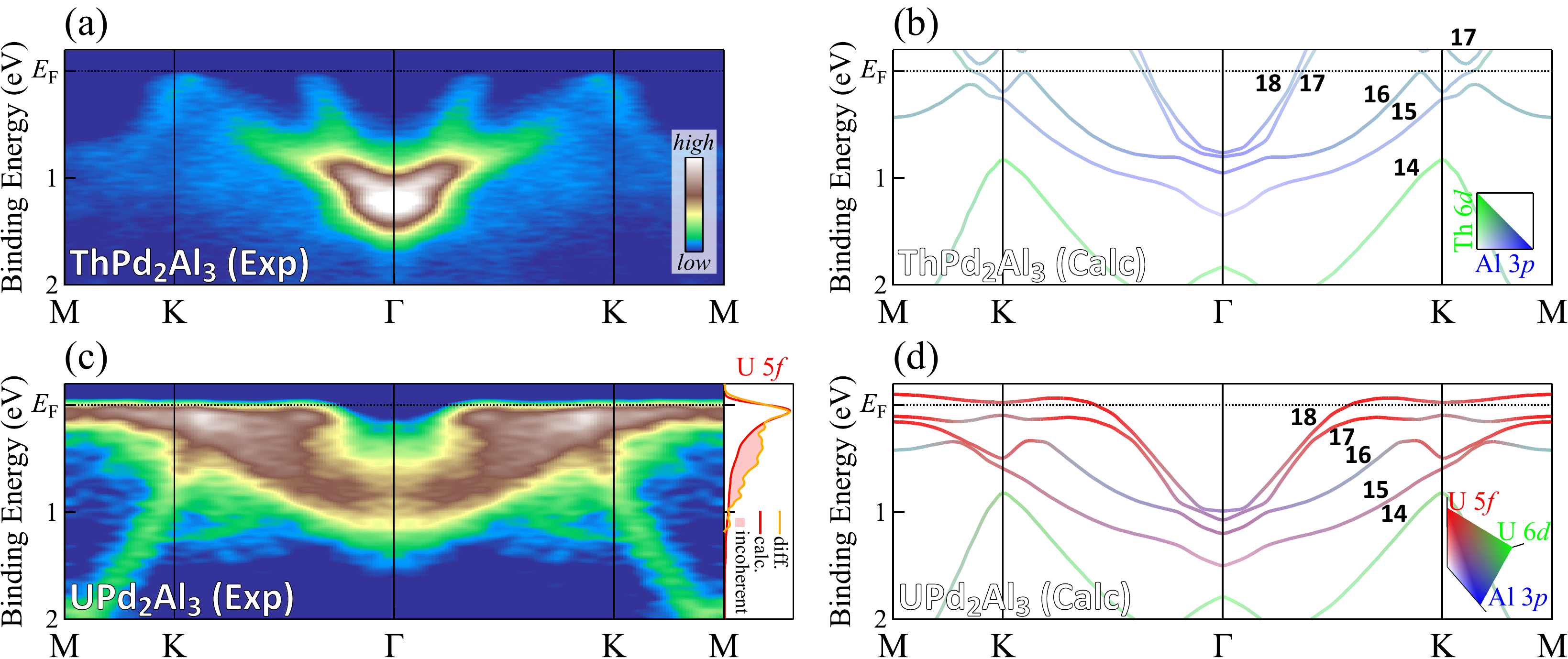}
	\caption{Comparison between the ARPES spectra of \ThPdAl and \UPdAl, and the result of the band-structure calculations.
	(a) ARPES spectra of \ThPdAl measured along the \Gm-\pnt{K}-\pnt{M} high-symmetry line at \hn{=660}.
	(b) Result of the band-structure calculation for \ThPdAl.
	The color coding of each band represents the contributions from the \orb{Th}{6d} and the \orb{Al}{3p} states.
	(c) ARPES spectra of \UPdAl measured along the \Gm-\pnt{K}-\pnt{M} high-symmetry line at \hn{=600}.
	Data adapted from Ref.~\cite{SF_review_JPSJ}.
	The \Uf difference spectrum obtained by resonant photoemission measurement \cite{U4d5fRPES} and the calculated \Uf DOS are	 shown in the right panel.
	(d) Result of the band-structure calculation for \UPdAl.
	The color coding of each band represents the contributions from the \Uf, \orb{U}{6d} and the \orb{Al}{3p} states.
}
	\label{ThU}
\end{figure*}
Next, we discuss the band structure and Fermi surface of \ThPdAl measured by ARPES.
Figure~\ref{arpes} summarizes the result of the ARPES study of \ThPdAl.
Figure~\ref{arpes} (a) shows the ARPES spectra of \ThPdAl along the \pnt{K}-\pnt{M}-\pnt{K}-\Gm-\pnt{M}-\Gm high-symmetry line measured at \hn{=660}.
The spectra consist of two different types of energy dispersions, namely very dispersive bands at $E_{\mathrm{F}} \lesssim E_{\mathrm{B}} \lesssim 2.5~\mathrm{eV}$ and less dispersive bands with enhanced intensities distributed at \EB{ \gtrsim 2.5}.
Since the photoionization cross section of the \Pdd states is more than 30 times enhanced than those of \orb{Al}{3s, 3p} and \orb{Th}{6d} states \cite{atomic}, the prominent dispersions at \EB{ \gtrsim 2.5} are the contributions from the \Pdd states.
Furthermore, the dispersive bands at $E_{\mathrm{F}} \lesssim E_{\mathrm{B}} \lesssim 2.5~\mathrm{eV}$ reflect the contributions mainly from the $\mathrm{Al}$ and $\mathrm{Th}$ states.

Figure~\ref{arpes} (b) shows the calculated band structure and the simulation of the ARPES spectra based on the band-structure calculation.
The solid lines and the density plot represent the calculated energy dispersions and the simulation, respectively.
The color coding of each band represents the contribution from the \Pdd states.
The experimental band structure is well-explained by the band-structure calculation.
For example, the parabolic dispersions centered at the $\mathrm{\Gamma}$ point that forms the Fermi surface are in good agreement with the calculated bands 16-18. 
There are also parabolic dispersions along the \pnt{K}-\pnt{M}-\pnt{K} high-symmetry line, which are also well-explained by the calculated band 15.
At higher binding energies, the nearly flat \Pdd bands in the experimental ARPES spectra agree with the bands 2-13 in the calculation.
Although the \Pdd bands are not clearly resolved one by one in the experimental spectra, the overall features agree very well between experimental ARPES spectra and the calculated band structure.
Note that the contribution from the \Pdd states is distributed in higher binding energies ($E_{\mathrm{B}} \gtrsim 2.5~\mathrm{eV}$), and the Fermi surface of \ThPdAl mainly consists of the $\mathrm{Al}$ and $\mathrm{Th}$ states.
This situation is significantly different from the case of \ThRuSi where the transition metal $d$ band is close to the Fermi energy, and the $d$ bands are hybridized with the \Uf states in \URuSi \cite{ThRu2Si2_ARPES,URu2Si2_3D}.
We further discuss the orbital character of the Fermi surface of \ThPdAl and \UPdAl in the later of this section.

Figure~\ref{arpes} (c) shows the experimental Fermi surface map of \ThPdAl, which was obtained by integrating the photoemission intensity over $100~\mathrm{meV}$ across \EF.
There is a hexagonal-shaped feature with enhanced intensity centered at the \Gm point.
In addition, the intensity at the \pnt{K} point is also enhanced and these points are connected to each other and form a very complex shape.
In Fig.~\ref{arpes} (d), we illustrate the calculated Fermi surface using solid curves, and we also demonstrate the simulation the Fermi surface map based on the band-structure calculation as a density plot.
The three dimensional shape of the calculated Fermi surfaces of \ThPdAl is also shown in Fig.~\ref{arpes} (e).
In the band-structure calculation, the bands 16-18 form Fermi surfaces.
Bands 16 and 18 form a hole-type Fermi surface around the \pnt{M} point and an electron-type Fermi surface around the \Gm point, respectively.
In contrast, band 17 forms a very complicated Fermi surface with a three-dimensional shape, but forms the electron-type Fermi surface with a star-like shape around the \Gm point.
Experimental and calculated Fermi surfaces agree very well although the features are broader in the experimental map compared to the calculated map.
Despite the fact that the details of the experimental Fermi surface are not very clear, the topology of the Fermi surface agrees with the result of the band-structure calculation.
Accordingly, the band structure and Fermi surface of \ThPdAl were well-explained by the band-structure calculation.

Next, we performed a comparison between the ARPES spectra of \ThPdAl and \UPdAl to reveal the contribution of the \Uf states in the band structure of \UPdAl, as shown in figure~\ref{ThU}.
Figures~\ref{ThU} (a) and (b) represent the experimental ARPES spectra of \ThPdAl and the corresponding calculated band structure, respectively.
The color coding of the calculated bands corresponds to the contributions from the \Thd and the \Alp states.
Note that the contribution from the \orb{Al}{3s} is almost negligible in these binding energies.
The experimental ARPES spectra are well-explained by the band-structure calculation, as discussed in Figs.~\ref{arpes}, and thus the orbital character of each band should also agree with the result of the band-structure calculation.
The calculation suggests that the Fermi surface of \ThPdAl mainly consists of the \Alp and \Thd states, but it has an enhanced \Alp character.
In particular, bands 17 and 18, which form the Fermi surfaces, have a dominant contribution from the \Alp states, and the overall good agreement between the experimental data and the respective calculations suggests that the Fermi surfaces should have an enhanced contribution from the \Alp states.

Figures~\ref{ThU} (c) and (d) show the experimental ARPES spectra of \UPdAl and the corresponding calculated band structure.
The ARPES spectra of \UPdAl were taken from Ref. \cite{SF_review_JPSJ}. 
The \Uf difference spectrum obtained by resonant photoemission measurement \cite{U4d5fRPES} and the calculated \Uf DOS are shown in the right panel of Fig.~\ref{ThU} (c).
The ARPES spectra were recorded at \hn{=600}, and the sample temperature was $20~\mathrm{K}$.
The experimental energy dispersions of \ThPdAl and \UPdAl are very similar to each other, but the intensity of the energy dispersions distributed at \EB{=E_{\rm{F}}-1.2} is enhanced in the spectra of \UPdAl.
Moreover, there exist very flat features in the vicinity of \EF, which represent the contributions from the \Uf states since the photoionization cross section of the \Uf states is dominant at the used photon energy.
The overall structure of the experimental spectra of \UPdAl are also essentially explained by the band-structure calculation although the detail of each dispersion was not resolved experimentally.
In particular, the very flat features at \EF originate from the renormalized \Uf bands due to the strong electron correlation effect.

The comparison between the ARPES spectra of \ThPdAl and \UPdAl indicates that the \Uf states are strongly hybridized with the non-$f$ dispersive bands in \ThPdAl which correspond to the calculated bands 15-18 of \ThPdAl.
These calculated bands have the enhanced contribution from the \Alp states, suggesting that the \Uf states are strongly hybridized with the \Alp states in \UPdAl.
The crystal structure of \UPdAl consists of alternating stacks of $\mathrm{U}$--$\mathrm{Pd}$ and $\mathrm{Al}$ layers along the $c$ axis, and the presence of the enhanced \Uf -- \Alp hybridization suggests that the \Uf states have a strong three-dimensional nature similar to the case of heavy fermion compound \URuSi \cite{URu2Si2_3D}.
As a result, the Fermi surface of \UPdAl should also have a strong three dimensional nature due to the enhanced \Uf--\Alp hybridization.
Previous ARPES studies of \UPdAl \cite{UM2Al3_ARPES,UPd2Al3_ARPES2} have shown a cylindrical Fermi surface along the \Gm -- \pnt{A} line.
The Fermi surface has an enhanced contribution from the \Uf states, but the present result suggests that it also has the enhanced contribution from the \Alp states.

Note that the \Uf states have a strong electron correlation effect, and the \Uf states in \UPdAl have incoherent peak distributed at approximately \EB{= 0.2-1} \cite{U4d5fRPES}.
Thus, the experimental bands around this binding energy should have the contribution from the incoherent component of the \Uf states.
This is consistent with the dual nature of the \Uf states in \UPdAl as proposed theoretically, where the heavy quasi-particle band is described by the effective renormalized theory while the high-energy structure is explained by the multiplet side bands arising from the Hund's rule \cite{UPd2Al3_Zwicknagl2,UPd2Al3_Zwicknagl3}.
Experimentally, there are not non-dispersive but dispersive bands in this binding energy region, and thus the energy dispersions in \EB{= 0.2-1} are hybridized with the incoherent \Uf states.
It is theoretically proposed that the localized multiplet bands are hybridized with dispersive non-$f$ bands \cite{UPd2Al3_Zwicknagl3}, and the experimental spectra are consistent with the theory.

\section{Conclusion}
In the present study, we investigated the electronic structure of \ThPdAl using photoelectron spectroscopy, and the results were compared with the band-structure calculation and the spectra of the isostructural heavy fermion superconductor \UPdAl.
The \Pdd states in the valence-band spectrum of \ThPdAl were found to be shallower than those in the valence-band spectrum of \UPdAl by approximately $200~\mathrm{meV}$, suggesting that the \Uf states are involved in the valence-band structure of \UPdAl, and are not impurity-like contributions.
The Fermi surface and the band structure of \ThPdAl obtained by ARPES were well explained by the band-structure calculation.
The electronic structure in the very vicinity of \EF is dominated by contributions from the \Alp and \Thd states with minor contributions from the \Pdd states.
The comparison between the ARPES spectra of \ThPdAl and \UPdAl suggests that the the electronic structure of \UPdAl in the very vicinity of \EF is dominated by the enhanced \Uf -- \Alp hybridization.
This indicates that the electronic structure of \UPdAl has a three-dimensional nature.

\nocite{LAPW}
\nocite{other}
\nocite{RelatElectron}
\nocite{APW}
\nocite{OAnderson}
\nocite{SPCD}
\nocite{tetra}

\begin{acknowledgments}
The experiment was performed under Proposal No. 2015B3820 at SPring-8 BL23SU.
The present work was financially supported by JSPS KAKENHI Grant Numbers JP26400374, JP16H01084, JP18K03553, and 	JP20KK0061.
\end{acknowledgments}

\bibliography{ThPd2Al3}

\end{document}